# Mechanical properties of boron phosphides

Vladimir L. Solozhenko [a,*] and Volodymyr Bushlya [b]

[a] *LSPM–CNRS, Université Paris Nord, 93430 Villetaneuse, France*
[b] *Division of Production and Materials Engineering, Lund University, 22100 Lund, Sweden*

Microstructure and mechanical properties of bulk polycristalline boron phosphides (cubic BP and rhombohedral $B_{12}P_2$) have been studied by scanning electron microscopy and micro- and nanoindentation. The obtained data on hardness, elastic properties and fracture toughness clearly indicate that both phosphides belong to a family of hard phases and can be considered as prospective binders for diamond and cubic boron nitride.

Cubic (*F*-43*m*) BP and rhombohedral (*R*-3*m*) $B_{12}P_2$ boron phosphides are refractory (melting temperatures at ambient pressure are 2840 K [1] and 2390 K [2]) and low-compressible (300-K bulk moduli are 174 GPa [3] and 192 GPa [4]) wide bandgap semiconductors that have attracted considerable attention due to their superior physical properties. Very recently two new methods of boron phosphides production have been developed i.e. self-propagating high-temperature synthesis [5,6] and ultra-fast mechanochemical synthesis [7]. Both methods are characterized by the simplicity of implementation, high efficiency and low production cost that together opens up broad perspectives for the large-scale production of high-purity micropowders of boron phosphides. According to the predictions made in the framework of thermodynamic model of hardness [8], both boron phosphides should exhibit Vickers hardness in the range of 30-35 GPa [9,10] comparable to that of polycrystalline boron carbide. In the present paper we report the experimental results on mechanical properties of boron phosphides.

**Experimental**

Polycrystalline powders of single-phase stoichiometric boron phosphides have been produced by self-propagating high-temperature synthesis according to the method described previously [5,6]. Well-sintered polycrystalline bulks of boron phosphides have been produced in a toroid-type apparatus with a specially designed high-temperature cell [11] by sintering of powder compacts at 5-7 GPa and temperature close to the melting point. According to X-ray diffraction study (TEXT 3000 Inel, CuKα1 radiation) the recovered bulks contain well-crystallized single-phase BP

[*] vladimir.solozhenko@univ-paris13.fr

($a$ = 4.535(1) Å) or $B_{12}P_2$ ($a$ = 5.992(4) Å, $c$ = 11.861(8) Å) with lattice parameters close to the literature data [12,13].

The recovered samples (cylinders 4-mm diameter and 3-mm height) were hot mounted in electrically conductive carbon-fiber reinforced resin, and were planar ground with diamond 500 grit and subsequently polished with 9-μm and 1-μm diamond suspensions. Mechanical polishing was followed by vibropolishing with 0.04-μm $SiO_2$ colloidal solution for 8 hours that ensured the minimal sample surface damage.

Microstructure of the polished samples have been studied by high-resolution scanning electron microscopy (SEM) using Hitachi SU8010 Cold Field Emission SEM microscope in secondary electron and backscatter modes.

Microhardness measurements have been performed using a Ernst Leitz Wetzlar indentation tester under loads ranging from 0.25 to 6.0 N and 15 seconds dwell time. At least, five indentations have been made at each load. The indentation imprints were measured with a Leica DMRME optical microscope under 1000× magnification in the phase contrast regime. Vickers hardness ($H_V$) was determined from the residual imprint upon indentation and was calculated following the standard definitions according to Eq. 1:

$$H_V = \frac{1.8544 \cdot P}{d^2} \quad (1)$$

where $P$ and $d$ are the applied load and residual imprint diagonal, respectively. The value of Knoop hardness ($H_K$) was determined by Eq. 2:

$$H_K = \frac{P}{0.070279 \cdot d^2} \quad (2)$$

where P is the applied load and $d$ is the length of a large diagonal of an imprint.

Nanoindentation study has been performed on Micro Materials NanoTest Vantage system with trigonal Berkovich diamond indenter (the tip radius of 120 nm). The maximal applied load was 500 mN. Loading at the rate of 0.5 mN/s was followed by a 10 s holding and unloading at the same rate. AFM microscope Dimension 3100 (Digital Instruments) in tapping mode was used on nanoindentation imprints for pile-up correction.

Evaluation of the hardness and elastic modulus was performed in accordance to the Oliver-Pharr method [14]. The nanohardness was determined by Eq. 3:

$$H_N = \frac{P_{max}}{A(h_c)} \quad (3)$$

where $P_{max}$ is the maximum applied load and $A(h_c)$ is the projected contact area. The area function $A(h_c)$ was calibrated on a standard fused silica reference sample. Correction of the area function for the pile-up effects was based on the imprint topography data obtained on the actual samples by atomic force microscopy. The elastic recovery was estimated as the ratio of elastic work to the total work of the indentation by Eq. 4:

$$R_W = \frac{W_e}{W_p + W_e} \times 100\% \tag{4}$$

where $W_p$ and $W_e$ are plastic and elastic works, respectively. Reduced modulus $E_r$ was determined from stiffness measurements that are governed by elastic properties of the sample and diamond indenter via Eq. 5:

$$E_r = \left(\frac{1-v_s^2}{E_s} + \frac{1-v_i^2}{E_i}\right)^{-1} \tag{5}$$

where $E_s$, $E_i$ are Young's moduli and $v_s$, $v_i$ are the Poisson's ratios of the sample and indenter, respectively. The elastic modulus of material can be calculated for known properties of diamond ($E_i = 1141$ GPa and $v_i = 0.07$ [14]) and Poisson's ratio of the sample. Using the relation (6) between Young's ($E$) and shear ($G$) moduli of an isotropic material

$$G = \frac{E}{2(1+v)} \tag{6}$$

the shear modulus can be evaluated for the known value of Poisson's ratio.

The indentation fracture toughness ($K_{IC}$) was studied with a THV-30MDX hardness tester using Vickers diamond indenter under loads ranging from 5 to 100 N. The lengths of radial cracks emanating from the imprint corners were measured in polarized light with Alicona InfiniteFocus 3D optical microscope under 1000× magnification. The value of $K_{IC}$ was determined in terms of the indentation load $P$ and the mean length (surface tip-to-tip length $2c$) of the radial cracks according to Eq. 7 [15]:

$$K_{IC} = x_v \cdot (E/H_V)^{0.5} (P/c^{1.5}) \tag{7}$$

where $x_v = 0.016(4)$, $E$ is Young's modulus and $H_V$ is load-independent Vickers hardness. Loads were restricted to values for which the minimum requirement to the relation of radial crack $c$ to half-diagonal $a = d/2$ was maintained at $c \geq 2a$.

**Results and Discussion**

**Boron phosphide**

Secondary electron image of the bulk BP shows polyhedral grains having dimensions of 2 to 8 μm that are characteristic for the entire sample (Fig. 1a). The observed residual porosity, primarily on triple joints, is associated with grain growth at sintering temperature (2330 K at 6.5 GPa) that is slightly below the melting point.

The measured Vickers hardness of boron phosphide decreases with the load and at 4 N reaches the asymptotic value $H_V = 29(2)$ GPa (Fig. 2a) that is in perfect agreement with the value previously calculated in the framework of the thermodynamic model of hardness [9].

The load dependence of the measured Knoop hardness is presented in Fig. 2b; the asymptotic hardness value of $H_K = 20(2)$ GPa is however rather low and demonstrates abnormal lag to the Vickers hardness value. This can be attributed to low fracture toughness (see below for details) and related cleavage of the material under Knoop indenter (see Inset in Fig. 2b).

From 15 independent nanoindentation experiments it was found that the measured nanohardness of boron phosphide decreases with peak indentation load from 31 GPa at 50 mN down to the asymptotic value of $H_N = 26(1)$ GPa at 500 mN. Fig. 3 shows a characteristic load-displacement curve. The elastic recovery of BP has been estimated by Eq. 4 as 55(3)% that is close to elastic recovery of single-crystal cubic BN (60% [16]). The Young's modulus of $E = 362(36)$ GPa was calculated by Eq. 5 using the experimental value $E_r = 280(19)$ GPa and Poisson's ratio $\nu = 0.15$ estimated from the experimental value of BP bulk modulus (174 GPa [3]) according to the relation between bulk and Young's moduli

$$B = \frac{E}{3(1-2\nu)} \tag{8}$$

Thus, the theoretically predicted value of BP Young's modulus (390 GPa) reported earlier [17] is somewhat overestimated. The shear modulus of boron phosphide was evaluated by Eq. 6 as $G = 157(16)$ GPa which is in between the theoretically predicted values i.e. 139 GPa [18] and 173 GPa [17].

Fracture toughness data for BP are presented in Fig. 4a. It can be seen that $K_{IC}$ value exhibits a strong load dependence initially decreasing down to $K_{IC} = 1.6$ MPa·m$^{0.5}$ at 20 N load, which is then followed by a monotonic increase up to 2.9 MPa·m$^{0.5}$ at 50 N. This change in behaviour can be attributed to fracture and spalling of the material by the indenter facets under loads higher than 20 N, as seen in Fig. 4c. At 20 N load the crack lengths (70–80 μm) are much longer than the grains, so the measured $K_{IC}$ value is characteristic of the bulk material as a whole.

**Boron subphosphide**

Fig. 1b shows the secondary electron image of bulk $B_{12}P_2$ fabricated at 5.2 GPa by crystallization from the melt at the temperature slightly above the melting point. The sample is homogeneous and has grain size of 0.2 to 1 μm. Similarly to BP, the residual porosity is also observed.

The measured Vickers hardness of boron subphosphide as a function of load is presented in Fig. 5a; the asymptotic value $H_V = 35(2)$ GPa is attained at 3 N. This value is in excellent agreement with $B_{12}P_2$ hardness calculated in the framework of the thermodynamic model of hardness [10], and is comparable to the hardness of commercial polycrystalline cubic boron nitride. Load dependence of Knoop hardness (Fig. 5b) is characterized by an asymptotic value of $H_K = 26(2)$ GPa which is achieved already at 2 N load.

Load-displacement curve of boron subphosphide is shown in Fig. 3. In the whole studied range of peak indentation loads (50-500 mN) the measured nanohardness is virtually constant and makes $H_N = 31.3(6)$ GPa. The elastic recovery of BP has been estimated as 56(3)% which is very close to that of boron phosphide.

Young's modulus $E = 454(23)$ GPa was calculated by Eq. 5 using the experimental value $E_r = 328(5)$ GPa (data from 15 independent nanoindentation experiments) and Poisson's ratio $\nu = 0.11$ estimated from relation (8) and bulk modulus of $B_{12}P_2$ (192 GPa [4]). Using the Eq. 6, the shear modulus of $B_{12}P_2$ was evaluated as $G = 205(11)$ GPa.

Similarly to boron phosphide, fracture toughness of $B_{12}P_2$ also exhibits a pronounced load dependence (Fig. 4b); the maximum value of $K_{IC} = 3.9$ MPa·m$^{0.5}$ is observed at 25 N load. The crack lengths at this load (40–50 μm) are significantly longer than the grains, so the $K_{IC}$ value is a characteristic of the bulk $B_{12}P_2$ material.

**Conclusions**

The data on mechanical and elastic properties of boron phosphides are summarized in the Table. Due to high hardness and elastic recovery as well as high thermal stability both boron phosphides offer promise as potential binders for diamond and cubic boron nitride.

**Acknowledgments**

The authors thank Dr. Vladimir A. Mukhanov for fabrication of boron phosphides' bulks and Prof. Jinming Zhou for nanoindentation study.  This work was financially supported by the European Union's Horizon 2020 Research and Innovation Programme under the Flintstone2020 project (grant agreement No 689279).

Table  Hardness, elastic moduli and fracture toughness of boron phosphides

|  | $H_V$ | $H_K$ | $H_N$ | $E$ | $G$ | $B$ | $K_{IC}$ (25 N) |
|---|---|---|---|---|---|---|---|
|  | GPa | | | | | | MPa·m$^{0.5}$ |
| BP | 34(2) | 20(2) | 26(1) | 362(36) | 157(16) | 174 [3] | 1.8(3) |
| B$_{12}$P$_2$ | 35(2) | 26(2) | 31.3(6) | 454(23) | 205(11) | 192 [4] | 3.9(8) |

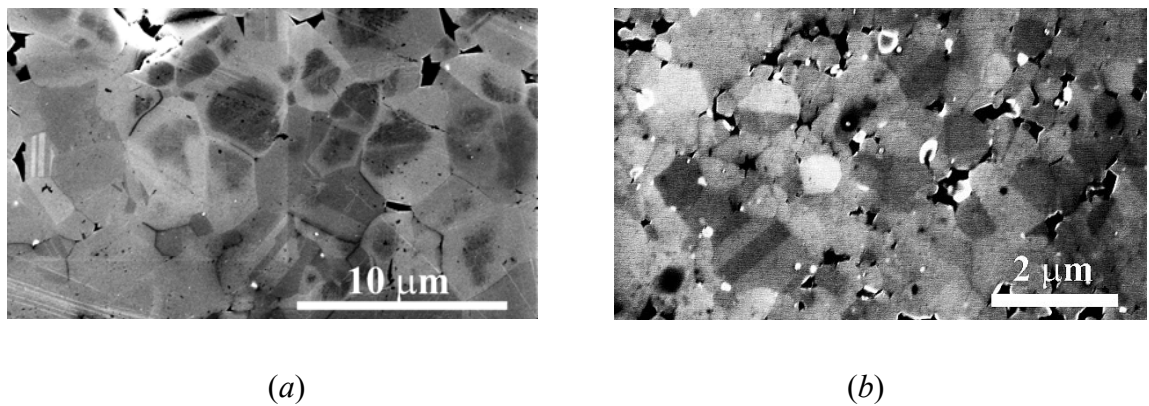

(*a*)                                             (*b*)

Fig. 1    Secondary electron images of the microstructure of BP (*a*) and $B_{12}P_2$ (*b*) bulks.

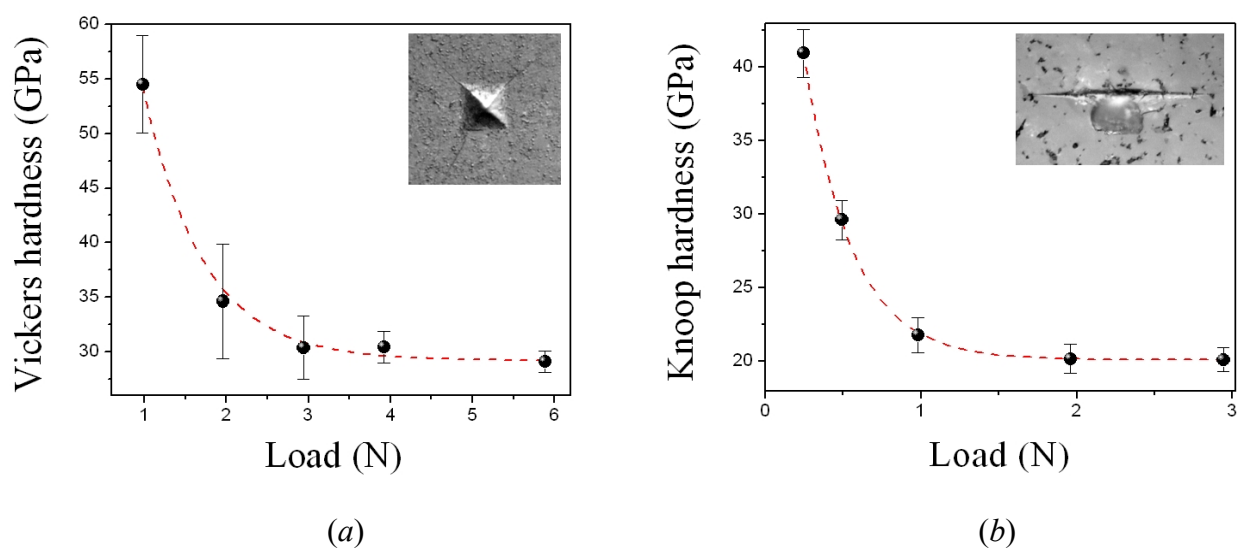

Fig. 2   Vickers (*a*) and Knoop (*b*) microhardness of bulk boron phosphide BP *vs* load. Insets: optical microscope images of the imprints formed by Vickers and Knoop indenters under 3 N loads.

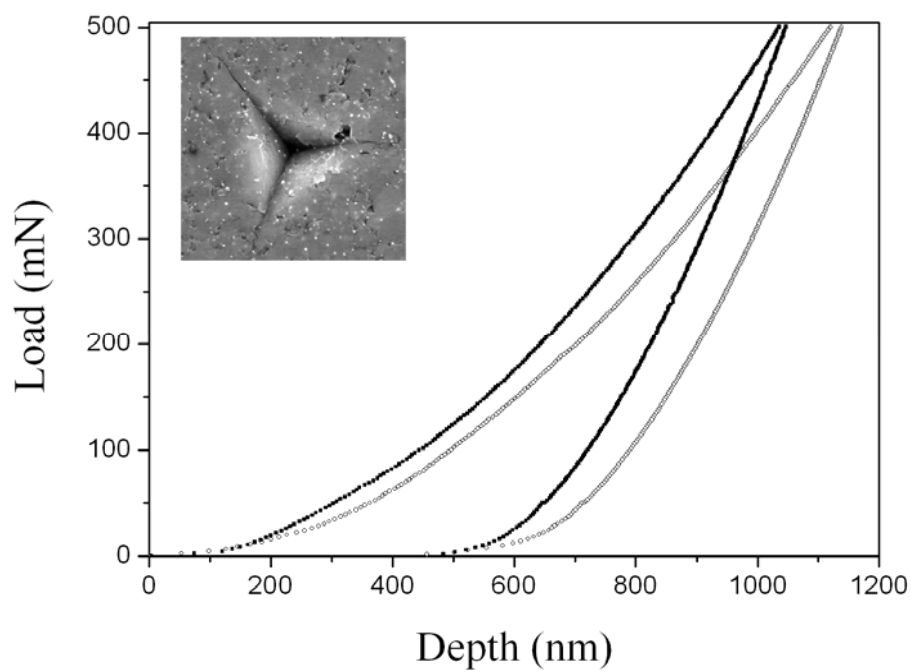

Fig. 3  Characteristic load-displacement curves for bulk boron phosphides BP (open symbols) and $B_{12}P_2$ (solid symbols). Inset: SEM image of the residual imprint formed by Berkovich indenter on the surface of $B_{12}P_2$ sample at 300 mN peak indentation load.

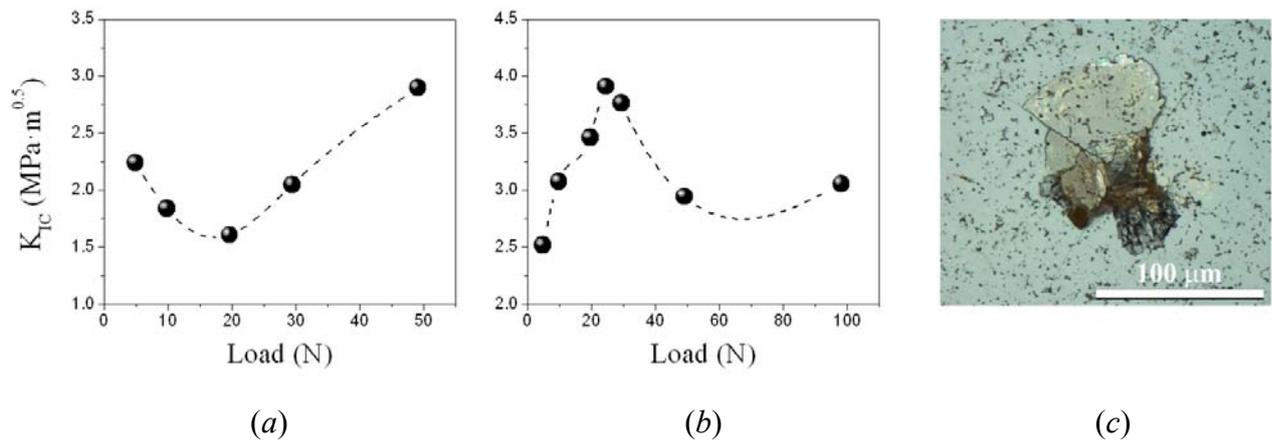

Fig. 4　Fracture toughness of BP (*a*) and $B_{12}P_2$ (*b*) boron phosphides. Optical microscope image of BP fracture and spalling at 30 N load (*c*).

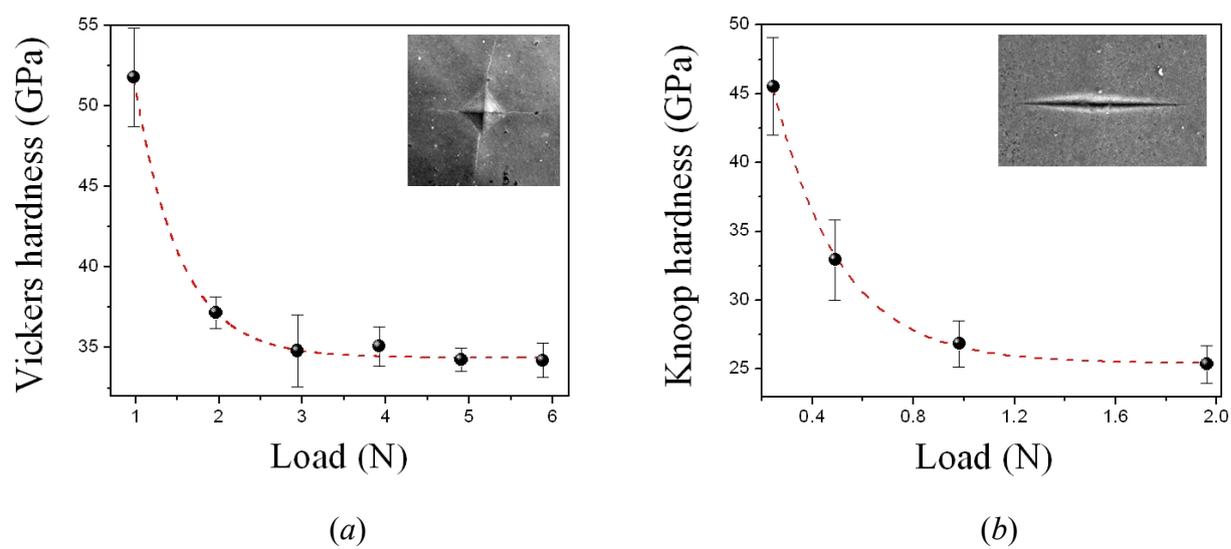

Fig. 5  Vickers (*a*) and Knoop (*b*) microhardness of bulk boron subphosphide $B_{12}P_2$ *vs* load. Insets: SEM images of the imprints formed by Vickers and Knoop indenters under loads of 3 N and 2 N, respectively.